\documentstyle[prb,aps, fleqn]{revtex}

\begin{document}
\title{ Magnetic properties of exactly solvable doubly
           decorated  Ising-Heisenberg  planar models}
\author{Jozef Stre\v{c}ka and Michal Ja\v{s}\v{c}ur \\
\normalsize Department of Theoretical Physics and Geophysics,
Faculty of Science,\\
\normalsize P. J. \v{S}af\'{a}rik University,
Moyzesova 16, 041  54 Ko\v{s}ice, Slovak Republic\\
\normalsize E-mail: jascur@kosice.upjs.sk\\
                    jozkos@pobox.sk }
\date{Submitted: \today}
\maketitle

\begin{abstract}
Applying the decoration-iteration procedure, we introduce a class
of exactly solvable doubly decorated planar models consisting both of the
Ising- and Heisenberg-type atoms.
Exact solutions for the ground state, phase diagrams and basic physical
quantities are derived and discussed. The detailed analysis of the
relevant quantities suggests the existence of an interesting
quantum antiferromagnetic phase in the system.
\end{abstract}
\section{Introduction}
The quantum Heisenberg model (QHM) \cite{[1]} and its simplified Ising version
\cite{[2]} remain to be one of the most actively studied subjects in
statistical mechanics.
In particular, the low-dimensional antiferromagnetic QHM has recently
attracted a lot of attention since it represents a very useful model for
the investigation of interesting quantum phenomena. Among the most fascinating
problems that are of current interest in this field one
should mention: the investigation
of Haldane gaps in the 1D antiferromagnetic QHM with integer spins \cite{[3]},
quantum phase transitions  \cite{[4]},  spin-Peierls instabilities
\cite{[5]},
dimerization and other related phenomena \cite{[6]},
quantum entanglement \cite{[7]}, magnetization plateau \cite{[8]}, and so on.
In addition to the above mentioned works,  a number of the studies has been
devoted to the investigation of the role of magnetic ordering in
high-$T_c$ superconducting cuprates
consisting of two-dimensional Cu-O networks \cite{[9]}.
In fact, the magnetic properties of
these  materials can be well-described by means of a spin-$1/2$
antiferromagnetic 2D QHM.

Despite of extensive studies, the QHM  has been exactly solved in one dimension
only \cite{[10]}, thus for  higher dimensions only more or less accurate
approximate methods are available \cite{[11]}.
In general, the main difficulties of a
rigorous treatment of the QHM are closely associated with the noncommutability
of the spin operators involving the Hamiltonian of the system.
This  principal mathematical intractability
of the QHM has  motivated us to
introduce a class of interesting exactly solvable models, consisting both of
the Ising- and Heisenberg-type atoms.
For this purpose, we utilize  the well known decoration-iteration
procedure that has been originally introduced by Syozi \cite{[12]}
and later remarkably generalized by Fisher \cite{[13]}.
In the spirit of the Syozi's and Fishers's papers,  we use in this work the
decoration procedure to put a couple of the
Heisenberg atoms on each bond of the regular
Ising lattice. In this way we obtain a doubly decorated model consisting of two
interpenetrating sublattices occupied by the Ising-
and Heisenberg-type atoms, respectively, and such a model can be of interest
both theoretically  and experimentally.
Theoretically, the
most outstanding feature of this model is the fact that it enables to
investigate, by an exact calculation, how the quantum Heisenberg atoms modify the
magnetic properties of the pure Ising systems.
On the other hand, from the experimental point of view, the model
can be inspiring in the preparation of new magnetic materials with similar
topological structure as the system under investigation.
Actually, some of recently synthesized compounds
represent a progressive step in this direction \cite{[14]}.

The outline of the present paper is as follows. In Sec. II,
the main points of the mathematical formulation of the decoration-iteration
procedure for the Ising-Heisenberg models are explained and
the exact equation for the phase diagrams and basic physical
quantities are derived.  The most interesting numerical results  are presented and discussed in detail
in Sec. III and finally, some concluding
remarks are given in Sec. IV.

\section{Formulation}
In this work we will study a spin-$1/2$  doubly
decorated Ising-Heisenberg model on planar lattices described
by the Hamiltonian:
\begin{eqnarray}
\hat {\cal H}_d = - \sum_{i, j} J \bigl [
         \Delta  (\hat S_i^x \hat S_j^x + \hat S_i^y \hat S_j^y)
                 + \hat S_i^z \hat S_j^z
                             \bigr ]
            - \sum_{k, l} J_1 \hat S_k^z \hat \mu_l^z,
\label{eq1}
\end{eqnarray}
where $\hat S_i^{\alpha}$ and $\hat \mu_l^z$ are the  well-known components
of spin-1/2 operators,
the parameters $J$ and $J_1$ represent the nearest-neighbor exchange
interactions that couple  the
pairs of Heisenberg atoms or Ising and Heisenberg atoms,
respectively (see Fig. 1).
As usual,  $\Delta$ describes  the spatial anisotropy
in the Heisenberg exchange interaction. In fact, this parameter
allows one to control  the behavior of the system between the
Ising-regime for $\Delta < 1$ (easy-axis like anisotropy)
 and the XY-regime for $\Delta > 1$ (easy-plane anisotropy).
In view of further
manipulations, it is useful to rewrite the total Hamiltonian of
the system in the form $\hat {\cal H}_d = \sum_{k} \hat {\cal H}_k$,
where the summation is carried out over all the pairs of Heisenberg
atoms. The bond Hamiltonian $\hat {\cal H}_k$ involves all the
interaction terms associated with $k$th couple of Heisenberg
atoms (see Fig. 1) and it is given by
\begin{eqnarray}
\hat {\cal H}_k = - J \bigl [
  \Delta (\hat S_{k1}^x \hat S_{k2}^x + \hat S_{k1}^y \hat S_{k2}^y)
         + \hat S_{k1}^z \hat S_{k2}^z \bigr ]
            - J_1 (\hat S_{k1}^z \hat \mu_{k1}^z
            + \hat S_{k2}^z \hat \mu_{k2}^z).
\label{eq2}
\end{eqnarray}
The most important point of our
treatment is the calculation of the partition function for the system under
investigation.
Taking into account the standard commutation relation
for the bond Hamiltonians (i. e., $[\hat {\cal H}_i,\; \hat {\cal H}_k ] = 0,
\quad i\ne k$), we can express the partition function ${\cal Z}_d$ of decorated
system in the form
\begin{eqnarray}
{\cal Z}_d &=& \mbox{Tr} \exp(- \beta \hat {\cal H}_d)
           = \mbox{Tr} \exp(- \beta \sum_{k=1}^{Nq/2} \hat {\cal H}_k)
\nonumber \\
          &=& \displaystyle \mbox{Tr}_{ \{ \mu \}} \prod_{k=1}^{Nq/2}
\displaystyle \mbox{Tr}_{S_{k1}} \displaystyle \mbox{Tr}_{S_{k2}}
\exp(- \beta  \hat {\cal H}_k), \quad \beta = 1/k_B T,
\label{eq3}
\end{eqnarray}
where $k_B$ being Boltzmann constant and $T$ the absolute
temperature. $N$ represents the total number of Ising atoms and
$q$ is the coordination
number of the original (undecorated) lattice.
The symbol $\mbox{Tr}_{ \{ \mu \}}$ means a trace over all
degrees of freedom of Ising spins and finally, $\mbox{Tr}_{S_{k1}}
\mbox{Tr}_{S_{k2}}$ denotes a trace over a couple of
Heisenberg spins residing  on the $k$th bond.
To proceed further, it is  useful to introduce the following extended
decoration-iteration transformation \cite{[12]},\cite{[13]}
\begin{eqnarray}
\mbox{Tr}_{S_{k1}} \mbox{Tr}_{S_{k2}} \exp(- \beta \hat {\cal H}_k) =
                   2 \exp(\beta J/4) \biggl \{
\cosh \Bigl[ \beta J_1 (\mu_{k1}^z &+& \mu_{k2}^z)/2 \Bigr] \nonumber \\
+ \exp(- \beta J/2) \cosh \Bigl[ \frac{\beta}{2}
\sqrt{J_1^2 (\mu_{k1}^z - \mu_{k2}^z)^2 + J^2 \Delta^2} \Bigr] \biggr \}
&=& A \exp(\beta R \mu_{k1}^z \mu_{k2}^z).
\label{eq4}
\end{eqnarray}
The unknown transformation parameters $A$ and $R$ can be easily
obtained following the standard procedures (see \cite{[12]} and references
therein), namely,
\begin{equation}
       A = 2 \exp(\beta J/4) (V_1 V_2 )^{1/2}, \quad
 \beta R = 2 \ln \Bigl( \frac{V_1}{V_2} \Bigr),
\label{eq5}
\end{equation}
where we have introduced the functions $V_1$ and $V_2$ as follows
\begin{eqnarray}
V_1 &=& \cosh(\beta J_1/2) + \exp(- \beta J/2) \cosh(\beta J \Delta/2 ),
\nonumber \\
V_2 &=& 1+ \exp(- \beta J/2) \cosh
        \Bigl( \frac{\beta}{2} \sqrt{J_1^2 + J^2 \Delta^2} \Bigr).
\label{eq6}
\end{eqnarray}
Now, after substituting Eq. (4) into Eq. (3) one obtains the equation
\begin{eqnarray}
{\cal Z}_d  = A^{Nq/2} {\cal Z}_0,
\label{eq7}
\end{eqnarray}
which relates the partition function of  doubly decorated Ising-Heisenberg
model (${\cal Z}_d $) to that of the original undecorated spin-1/2 Ising
model (${\cal Z}_0$.) From this simple relation, we can directly
calculate some physical quantities
(for example, the free and internal energy or  specific heat)
 on the basis of well-known thermodynamic relations. However, to understand the
behavior of the system, we have to analyze also some other quantities (for
instance the magnetization and pair-correlation functions) that cannot be
obtained from the partition function in a straightforward manner.
Fortunately, we can avoid this complication by exploiting
the following exact identities
\begin{equation}
\langle f_1 (\hat \mu_i^z,\hat  \mu_j^z, ... ,\hat \mu_k^z) \rangle_d =
\langle f_1 (\hat \mu_i^z,\hat  \mu_j^z, ... ,\hat \mu_k^z) \rangle_0,
\label{eq8}
\end{equation}
\begin{equation}
\langle f_2 (\hat S_{k1}^{\alpha}, \hat S_{k2}^{\gamma},
        \hat \mu_{k1}^z,  \hat \mu_{k2}^z) \rangle_d =
\biggl \langle \frac{\displaystyle \mbox{Tr}_{S_{k1}} \mbox{Tr}_{S_{k2}}
f_2 ( \hat S_{k1}^{\alpha},  \hat S_{k2}^{\gamma},  \hat \mu_{k1}^z,  \hat \mu_{k2}^z)
                         \exp(- \beta  \hat {\cal H}_k)}
     {\displaystyle \mbox{Tr}_{S_{k1}} \mbox{Tr}_{S_{k2}}
                   \exp(- \beta  \hat {\cal H}_k)}
\biggr \rangle_d,
\label{eq9}
\end{equation}
from which the relevant quantities can be calculated. In the above equations,
$f_1$ represents a function depending only on the Ising
spin variables and $f_2$ denotes a function which depends on
the spin variables located on the $k$th bond only. The superscripts
$\alpha$ and $\gamma$ denote $x, y$ or $z$ components of
the spin operators and
the symbols $\langle ... \rangle_d$ and $\langle ... \rangle_0$ mean the
standard ensemble average related to the decorated and original
lattice, respectively. Now,  applying one of the
standard methods \cite{[15]}, we simply derive the following equations for
sublattice magnetization
\begin{equation}
m_A^z \equiv
\frac12 (\langle \hat \mu_{k1} + \hat \mu_{k2} \rangle_d) =
\frac12 (\langle \hat \mu_{k1} + \hat \mu_{k2} \rangle_0) = m_0,
\label{eq10}
\end{equation}
\begin{equation}
m_B^z \equiv \frac12 \langle \hat S_{k1}^z + \hat S_{k2}^z \rangle_d =
8(\langle \hat \mu_{k1} \rangle_d + \langle \hat \mu_{k2} \rangle_d )K_0
\label{eq11}
\end{equation}
\begin{equation}
m_B^x \equiv \frac12 \langle \hat S_{k1}^x + \hat S_{k2}^x \rangle_d = 0,
\label{eq12}
\end{equation}
\begin{equation}
m_B^y \equiv \frac12 \langle \hat S_{k1}^y + \hat S_{k2}^y \rangle_d = 0,
\label{eq13}
\end{equation}
where  $m_0$ is the magnetization per one site of the original lattice and the
coefficient $K_0$ is given in the Appendix.
Similarly, the various pair correlations can be expressed with the
help of Eqs. (\ref{eq8}) and (\ref{eq9}) in the following simple form:
\begin{equation}
q_{ii}^{zz} \equiv \langle \hat \mu_{k1}^z \hat \mu_{k2}^z \rangle_d =
      \langle \hat \mu_{k1}^z \hat \mu_{k2}^z \rangle_0 \equiv \varepsilon,
\label{eq14}
\end{equation}
\begin{equation}
q_{hh}^{xx} \equiv \langle \hat S_{k1}^x \hat S_{k2}^x \rangle_d =
        K_1 + K_2 + 4q_{ii}^{zz} (K_1 - K_2),
\label{eq15}
\end{equation}
\begin{equation}
q_{hh}^{yy} \equiv  \langle \hat S_{k1}^y \hat S_{k2}^y \rangle_d
= q_{hh}^{xx},
\label{eq16}
\end{equation}
\begin{equation}
  q_{hh}^{zz} \equiv \langle \hat S_{k1}^z \hat S_{k2}^z \rangle_d
      = K_3 + K_4 + 4q_{ii}^{zz} (K_3 - K_4),
\label{eq17}
\end{equation}
\begin{equation}
q_{ih}^{zz} \equiv \frac12 \langle \hat S_{k1}^z \hat \mu_{k1}^z
                          + \hat S_{k2}^z \hat \mu_{k2}^z \rangle
         = K_0 + K_5 + 4q_{ii}^{zz}(K_0 - K_5).
\label{eq18}
\end{equation}
Here, $\varepsilon \equiv\langle \hat \mu_{k1}^z \hat \mu_{k2}^z \rangle_0$
denotes the nearest neighbor correlation of the original lattice that is well
known and the coefficients $K_0-K_5$ are listed in the Appendix.

Finally, the internal energy and specific heat of the system can be also easily
calculated from the relations
\begin{equation}
 U_d = - \frac{Nq}{2}\Biggl[ J\Delta (  q_{hh}^{xx} +   q_{hh}^{yy}) +
  Jq_{hh}^{zz}  + 2J_1   q_{ih}^{zz}
  \Biggr],
\label{eq19}
\end{equation}
\begin{equation}
C_d = \partial  U_d / \partial T.
\label{eq20}
\end{equation}

\section{Numerical Results and Discussion}
In this section we will show the most interesting numerical
results of the system under investigation. For the sake of
simplicity, we restrict our attention to case of the doubly decorated
square lattice (see Fig. 1) in which all characteristic properties can be
illustrated.

Before discussing the results, it is worth
noticing that the phase diagrams for the ferromagnetic $(J >
0, J_1 > 0)$ and ferrimagnetic $(J > 0, J_1 < 0)$ case will
be the same, since the relevant equation for the critical
temperature is invariant under the transformation
$J_1 \longleftrightarrow - J_1$.
On the other hand, the antiferromagnetic system $(J < 0$ and
arbitrary $J_1)$ exhibits many different features  and will
be discussed in a separate work.

In order to find possible ground state phases and to investigate their
properties, we have to analyze the internal energy,
magnetization and correlation functions at $T = 0$.
Depending on whether the  anisotropy parameter $\Delta$
is less, equal or greater then the boundary value
$\Delta_c = \sqrt{2|J_1|/J + 1}$ one finds  three different
regions. Namely,

i) for $\Delta< \Delta_c$
\begin{eqnarray}
\nonumber
&& U_d = -\frac{Nq}{8}\Bigl(2|J_1| + J\Bigr),\\
&&q_{ii}^{zz} = q_{hh}^{zz}  = 0.25, \quad q_{ih}^{zz} = \pm 0.25,
\quad q_{hh}^{xx}  = q_{hh}^{yy} = 0.0
\label{eq21}
\end{eqnarray}

ii) for $\Delta >  \Delta_c$
\begin{eqnarray}
\nonumber
&& U_d = -\frac{Nq}{8}\Bigl(-J + 2\sqrt{J_1^2 + (J\Delta)^2}\Bigr),\\
&&q_{ii}^{zz} = q_{hh}^{zz}  = -0.25,
\quad q_{ih}^{zz} = \pm \frac{J_1}{4\sqrt{J_1^2 +
(J\Delta)^2}},\nonumber \\
&& q_{hh}^{xx}  = q_{hh}^{yy} = \frac{J\Delta}{4\sqrt{J_1^2 + (J\Delta)^2}}
\label{eq22}
\end{eqnarray}

iii) for $\Delta = \Delta_c$
\begin{eqnarray}
\nonumber
&& U_d = -\frac{Nq}{8}\Bigl(2|J_1| + J\Bigr),\\
&&q_{ii}^{zz} = q_{hh}^{zz}  = 0.0,
\quad q_{ih}^{zz} = \pm \frac{2|J_1| + J}{8( |J_1| + J )},
\quad q_{hh}^{xx}  = q_{hh}^{yy} = \frac{\sqrt{J(|J_1| + J)}}{8( |J_1| + J) }
\label{eq23}
\end{eqnarray}
where the  plus or minus sign one applies for the ferromagnetic
or ferrimagnetic case, respectively. From these relations,
we have obtained the ground-state phase diagram in the
$|J_1|- \Delta$ space which is depicted in Fig. 2. Taking
into account Eqs. (\ref{eq21})-(\ref{eq23}), one easily identifies the standard
ferromagnetic (ferrimagnetic) phase (FP) for $\Delta <
\Delta_c$. However, for $\Delta > \Delta_c$
an unexpected quantum phase occurs in the system.
This phase (to be referred to as a quantum antiferromagnetic phase
(QAP)) requires more detailed description since it differs from
standard phases in the pure Ising or Heisenberg models and, as far as we know,
such a phase has not been described in the literature before.
In fact, from the relevant equations we find that in
the QAP  the Ising spins (that are the
nearest neighbors on the original lattice) are aligned
antiparallel with respect to each other.  Consequently,  we
have the classical N\'eel long-range ordering on the Ising sublattice
with $m_A = 0.0$ and $q^{zz}_{ii} = - 0.25$. On the other hand, the
nearest-neighboring Heisenberg spins create dimers,
thus we also have $m_B = 0.0$ and $q_{hh}^{zz}= - 0.25$.
However, the alignment of these dimers with respect to their
nearest-neighboring Ising spins is random, hence,
the relevant correlation function
$q^{zz}_{ih}$ does not reach its saturated value ($\pm 0.25$).
Moreover, one easily observes that the degree of randomness increases with the
increasing in anisotropy $\Delta$.
This behavior apparently appears due to the competition between strong
in-plane anisotropy ($\Delta$) that supports short-range ordering
and the exchange interactions ($J,\; J_1$) preferring the
long-range ordering along the easy axis.
It is also clear from the aforementioned arguments
that despite of  some disorder introduced
by random orientation of the dimers, the QAP will exhibit the long-range
N\'eel ordering captured to the Ising spins.
Thus, one can expect the appearance of the second-order
phase transition in the system at finite temperatures, even
for very strong values of the anisotropy parameter $\Delta$.
Nevertheless, the QAP also differs from the standard
N\'eel phase (both the classical and quantum one) due to the
ferromagnetic in-plane short-range order of the Heisenberg spins.
Another unusual feature of the QAP is the perfect
antiparallel alignment of the Ising spins (that are not directly coupled
via exchange), despite of some disorder present
between the nearest-neighboring Ising and Heisenberg atoms.
Furthermore, we would like to emphasize that the existence of the QAP
by itself is very surprising, since we have the system with the
ferromagnetic exchange interactions $J$ and $J_1$, only.

Finally, one should
notice that the FP and QAP are separated by the first-order
phase transition line that is given by the condition $\Delta = \Delta_c =
\sqrt{2 |J_1|/J + 1}$. At arbitrary point of this line there
coexist two of the above mentioned phases (FP, QAP) together with
a disordered phase (DP) in which we have $m^z_A = m^z_B =
q^{zz}_{ii} = q^{zz}_{hh} = 0.0$ and the nonvanishing
short-range ordering both in the $xy$ plane and  along the easy axis
$(q^{xx}_{hh} = q^{yy}_{hh} \not = 0, \quad
q^{zz}_{ih} \not = 0)$. The coexistence of these three
phases follows from the fact that the relevant ground-state energies take
the same value (in fact,
$ \lim_{\Delta \to \Delta_c^{-}}  U_d(\mbox{FP})
= \lim_{\Delta \to \Delta_c^{+}}  U_d (\mbox{QAP}) =
U_d(\mbox{DP})_{\Delta = \Delta_c}$).

Next, in order to demonstrate the overall
dependences of the correlation functions on the anisotropy
parameter $\Delta$, we have depicted in Fig. 3 the relevant
pair correlations for  the case  $J_1/J = 1.0$. In agreement with the arguments
given above, we find
the FP for $\Delta < \sqrt{3}$ and the QAP for $\Delta >
\sqrt{3}$. Moreover, Fig. 3 indicates that for
$\Delta > \sqrt{3}$, the correlations $q^{zz}_{ih}$ decreases with the
increasing $\Delta$ and vanishes in the limit of $\Delta \to \infty $.
On the other hand, the correlation $q^{xx}_{hh}$  increases with
the parameter $\Delta$ and approaches
its saturation value for $\Delta \to \infty$.
Contrary to this behavior, the
correlations $q^{zz}_{ii}$ and $q^{zz}_{hh}$ take the
saturation value independently of $\Delta$, excepting the special point
$\Delta = \Delta_c$ where they jump to zero.
Thus, the
antiparallel orientation of the relevant Ising and Heisenberg
spin pairs is not affected at all even by the very strong anisotropy.
Finally, to complete the ground-state analysis, we have
shown in Fig. 4 the internal energy of different phases as a
function of the anisotropy parameter $\Delta$ for $|J_1|/J =
1.0$. In this figure, the full and dashed lines represent
the stable and unstable parts of the relevant energies,
respectively. The
black point is the energy of the DP. This dependence
apparently supports our previous statements and clearly
illustrates the occurrence of the first-order phase transition
at $\Delta = \Delta_c = \sqrt{3}$.

Now, let us proceed to study the finite-temperature
phase diagrams for $q = 4$ that can be easily obtained by
putting $\beta_c R = \pm 2 \ln(1 + \sqrt{2})$ into Eq. (\ref{eq5}).
Solving the relevant equation numerically for some characteristic values of
$J_1$, we have obtained the phase
boundaries in the $\Delta - T_c$ plane that are plotted in Fig. 5.
 As one can see, the critical
temperature decreases gradually from its Ising value at $\Delta = 0$ and
vanishes for $\Delta = \Delta_c = \sqrt{2 |J_1|/J + 1}$. On the other
hand, for $\Delta > \Delta_c$ the transition temperature at first
rapidly  increases, then passes through a local maximum value and
finally tends to zero for $\Delta \to \infty$.
It is clear that for $\Delta < \Delta_c $
the phase boundary separates the FP and the DP, and similarly for
 $\Delta > \Delta_c$ separates QAP and DP.
As one can expect,  the relevant thermal phase transition is of the second
order and  belongs to the same
universality class as that of the usual 2D Ising model. Of course,
the thermal variations of physical quantities can differ from the
standard behaviors in the Ising model.
To illustrate the case, we have shown in Fig. 6 the temperature dependences of
the specific heat for $J_1/J = 0.5$. As we can see, in the isotropic case
($\Delta = 1.0 $) we have the standard dependence usually observed in the
Ising models. On the other hand,
for the values of the exchange anisotropy close
to the critical value ($\Delta_c = \sqrt{2}$), the specific heat may in
addition to the familiar Schottky-type maximum exhibit another maxima.
These maxima appear equally bellow (see the case $\Delta = 1.3$)
and above (see the insert in Fig. 6) the critical temperature,
as a consequence of the thermal excitations of the Heisenberg spins
that basically influence the ordering in the system.
The influence of the thermal excitations on the behavior of the system is
really of great importance and can be understood from the temperature
dependences of the correlation functions. For this purpose, we have plotted in
Fig. 7, 8 and 9 thermal variations of the correlations $q_{ih}^{zz}$,
$q_{hh}^{zz}$ and $q_{hh}^{xx}$ for the same values of $J_1$ and $\Delta$ as in
Fig. 6. It is clear from these figures (see the case $\Delta = 1.3$) that
the occurrence of the maximum bellow the critical temperature require
to satisfy two conditions:
i) the relevant correlations must increase (decrease) rapidly enough with the
increasing temperature in the low-temperature region (for
example, in the case of $\Delta = 1.0$ the additional maximum does not appear
due to the relatively slow excitation process in comparison with the case of
$\Delta = 1.3$).
ii) the critical temperature of the system must be relatively
high (for instance, in the case of $\Delta = 1.4$ we have
$k_B T_c/J \approx 0.0075$ that is very low for
the occurrence of the maximum below $T_c$, in spite of the very strong thermal
excitations).

In addition to this behavior, the double-peak specific heat curve
can be also observed here. The origin of the relevant maxima
above the critical temperature can be understood from the
thermal dependences of the correlation functions, as it is clearly
displayed in Fig. 6-9 for $\Delta = 1.42$. As one can see, the
correlation function $q_{ih}^{zz}$ rapidly increases in the relevant
region, though the short-range ordering of Heisenberg atoms in the $xy$
plane as well as in the $z$ direction is rapidly destroyed.
Very similar behavior
appears above $T_c$ also for $\Delta<\Delta_c$ excepting the fact
that in this case the correlations in the $xy$ plane abruptly
increases, although all the other correlations rapidly decreases as it is
apparent from Figs. 7-9 for the case of $\Delta = 1.4$.
Hence, the appearance of the multiple peaks in the specific heat
curve arises due to the relevant thermally induced
short-range ordering or disordering in the system. Finally, one should also mention that all pair-correlation
function exhibit at
critical temperature weak energy-type singularity known from the usual Ising
models.

Although, the numerical calculation have been presented
for the square lattice $q = 4$, we can on the basis of our formulation draw
some general conclusions about the behavior of the doubly
decorated Ising-Heisenberg systems. However,
one should emphasize that our next statements
do not concern the one-dimensional case that exhibits quite different behavior,
as result of the fact that no long-range order is possible in the system
at nonzero temperatures.

Firstly, it is clear from Eqs. (\ref{eq21})-(\ref{eq23}) that
the ground-state phase diagram does not depend
on the coordination number and spatial dimensionality of the
system.
This implies that the value of $\Delta_c$ is also
independent of the coordination number and  dimension, although our
preliminary investigation of the other systems has revealed that it depends on the
spin value of the Heisenberg atoms.
In general one can say that the QAP exists in many two-  and
three-dimensional lattices in the region of $\Delta>\Delta_c$. However, it is
necessary to emphasize the fact, that the above statement is valid only in the
case when the N\'eel order is possible on the relevant original lattice.
Apparently, this not the case, for example, of the doubly decorated triangular
or Kagom\'e lattices for which much more complicated phases will occur
for $\Delta>\Delta_c$.

It is also interesting to note that the QAP, as well as the standard FP
may exist at finite temperatures and the  temperature region
of their stability  is clearly enlarged with the increasing coordination
number and dimensionality of the system. This is a consequence of the fact
that the critical temperature increases with the increasing coordination number
and dimensionality of the system.
Of course, the thermal fluctuations gradually destroy the long-range
order (both for the FP and QAP)
in the system and if the temperature reaches its
critical value, then the system undergoes the second-order phase
transition sharing the same universality class as the standard spin-1/2 Ising
planar model.

\section{Conclusion}
In this work we have studied the doubly decorated
Ising-Heisenberg model on planar lattices. Applying the
standard decoration-iteration transformation, we have
obtained the simple relation between the partition function
of the decorated model and its corresponding standard
spin-1/2 Ising model. On the basis of this
relation, we have derived
the exact results for basic physical quantities that have been subsequently
discussed  for $J>0$ in Sec. III.

In our opinion, the prediction of the QAP is the most important finding of this
work and it illustrates how
the magnetic properties of the pure Ising systems can be
essentially modified by
introducing the quantum Heisenberg atoms on the bonds of the original lattice.
The origin of the QAP consists in the quantum
fluctuations arising in the system, hence the QAP itself provides a clear
manifestation of the quantum phenomena in the macroscopic scale.

Although at the present time we are not
aware of any experimental system which can be directly described by
the considered doubly decorated Ising-Heisenberg model,
we hope that the recent progress in molecular
engineering will result in the preparation of such materials.
In fact, recently synthesized compound
K$_3$Co$^{\mbox{III}}$(CN)$_6$.2Rh$_2^{\mbox{II}}$(CH$_3$COO)$_4$
(see Ref. [16]),
which has the magnetic structure of the doubly decorated square
lattice (Fig. 1), seems to be the most promising from
this point of view. Unfortunately, the Co$^{3+}$ ions (located
at the corners of each square) are in this compound due to the
very strong ligand field of the cyanide groups in the low-spin
state (i.e. they are diamagnetic). Nevertheless, tuning the
ligand field around the Co$^{3+}$ ions by the choice of other
ligands, represents a possible way how to prepare the compounds with
high-spin Co$^{3+}$ paramagnetic ions. Another possibility to
obtain the compounds of desired magnetic structure consists in the chemical
replacement of the Co$^{3+}$ ions by other transition metal
ions, such as Fe$^{3+}$ and Cr$^{3+}$ ions. Because  the
corner atoms of each square would possess spin 1/2 (in case of  Fe$^{3+}$
ion) or 3/2 (for  Cr$^{3+}$ ion), the present theory
could be applied to describe the behavior of these materials.
Since the most of the real materials have $\Delta \approx 1$,
it is also worth noticing that in the case of very weak exchange interaction $J_1$
($|J_1|<<J$), the QAP appears in our system near $\Delta \approx 1$
regardless of the coordination number and spatial
dimensionality. This supports our hope that the experimental confirmation of
this phase would be possible in some materials.

Finally, we would like to remark that the present formalism can be extended to
investigate many other interesting systems. Indeed, we have succeeded
in solving some interesting generalization of the system studied in this paper
and we have found very rich and interesting
behavior that will be discussed in forthcoming works.

Acknowledgment: This work has been supported by the Ministry of
Education of Slovak Republic under VEGA grant No. 1/9034/02.
\vspace*{1cm}

{\Large \bf Appendix}
\begin{eqnarray}
\nonumber
K_0 &=& \frac18 \frac{\sinh(\beta J_1 /2)}
{\cosh(\beta J_1 /2) + \exp(- \beta J /2) \cosh(\beta J \Delta/2)},
\nonumber
\\
K_1 &=& \frac18 \frac{\sinh(\beta J \Delta /2)}
  {\cosh(\beta J \Delta /2) + \exp(\beta J/2) \cosh(\beta J_1/2)},
\nonumber
\\
K_2 &=& \frac18  \frac{J \Delta}{\sqrt{J_1^2 + J^2 \Delta^2}}
  \frac{\sinh \Bigl(
\displaystyle \beta \sqrt{J_1^2 + J^2 \Delta^2}/2 \Bigr)}
   {\cosh \Bigl(
\displaystyle \beta \sqrt{J_1^2 + J^2 \Delta^2}/2 \Bigr)
                + \exp(\beta J/2)},
\nonumber
\\
K_3 &=&
\frac18 \frac{\cosh(\beta J_1/2) - \exp(- \beta J/2) \cosh(\beta J \Delta /2)}
             {\cosh(\beta J_1/2) + \exp(- \beta J/2) \cosh(\beta J \Delta
             /2)},\nonumber
\\
K_4 &=& \frac18 \frac{\exp(\beta J/2) -
         \cosh \Bigl(
\displaystyle \beta \sqrt{J_1^2 + J^2 \Delta^2}/2 \Bigr)}
             {\exp(\beta J/2) +
         \cosh\Bigl(
\displaystyle \beta \sqrt{J_1^2 + J^2 \Delta^2}/2 \Bigr)},
\nonumber\\
K_5 &=& \frac18  \frac{J_1}{\sqrt{J_1^2 + J^2 \Delta^2}}
         \frac{\sinh \Bigl( \beta \sqrt{J_1^2 + J^2 \Delta^2}/2
                     \Bigr)}
              {\cosh \Bigl( \beta \sqrt{J_1^2 + J^2 \Delta^2}/2
                     \Bigr) + \exp(\beta J/2)}.
\nonumber
\end{eqnarray}
The detailed derivation of the coefficients given above requires
lengthy calculation, thus is not presented here but it can be
obtained by the authors on the request.

\newpage
\begin{Large}
{\bf Figure captions}
\end{Large}
\begin{itemize}
\item [Fig. 1.]
The fragment of the doubly decorated square lattice.
The black (grey) circles denote the
positions of the Ising (Heisenberg) atoms, respectively. The
ellipse demarcates a typical bond described by the Hamiltonian
$\hat {\cal H}_k$ (see Eq. (2)).
\item [Fig. 2.]
Ground-state phase diagram of the doubly decorated Ising-Heisenberg model.
The full lines represents the line of first-order phase transitions that
separates the ferromagnetic or ferrimagnetic
phase (FP) from the quantum antiferromagnetic phase
(QAP)
\item [Fig. 3.]
Dependences of the pair-correlation functions at the frond state
on the anisotropy parameter $\Delta$ for the doubly decorated square
lattice ($q=4$) and $|J_1|/J = 1.0$.
\item [Fig. 4.]
Ground-state energy $U_d$ v.s. $\Delta$ for $q=4$ and $|J_1|/J = 1.0$.
The full and dashed lines represent, respectively,
the stable and unstable parts of the energies of relevant phases.
The black circle denotes the ground-state energy of the disordered phase (DP).
\item [Fig. 5.]
Phase boundaries in the $\Delta-T_c$ plane for  the doubly decorated
 square Ising-Heisenberg lattice ($q=4$) when the exchange interaction
$J_1$ is changed. FP, QAP and DP denote the ferromagnetic (or ferrimagnetic),
quantum antiferromagnetic and disordered phase, respectively.
\item [Fig. 6.]
Temperature dependences of the reduced specific heat of the doubly decorated
square lattice ($q=4$) for $J_1/J = 0.5$ and different values of the
anisotropy parameter $\Delta$.
The dashed (full) lines represent the cases corresponding to the
FP (QAP) ground-state phases, respectively.
 The insert shows the detail of the behavior
when the anisotropy parameter $\Delta$ takes the value close to the boundary
value $\Delta_c = \sqrt{2}$.
\item [Fig. 7.]
Temperature variations of the correlation function $q_{ih}^{zz}$ between
Ising and Heisenberg atoms for $q = 4$, $J_1/J = 0.5$ and different values of the
anisotropy parameter $\Delta$.
\item [Fig. 8.]
The same as in Fig. 7 but for the correlation $q_{hh}^{zz}$
between nearest-neighboring Heisenberg atoms.
\item [Fig. 9.]
The same as in Fig. 7 but for the correlation $q_{hh}^{xx}=q_{hh}^{yy}$
between nearest-neighboring Heisenberg atoms.
\end{itemize}
\end{document}